\documentclass[floatfix,
reprint,
superscriptaddress,
 amsmath,amssymb,
 aps,
pra,
]{revtex4-2}

\usepackage{gensymb}
\usepackage{graphicx}
\usepackage{dcolumn}
\usepackage{bm}
\usepackage[utf8]{inputenc}

\usepackage{graphicx}
\usepackage{physunits}
\usepackage{xcolor}
\usepackage{amsmath}
\usepackage{braket}
\usepackage{multirow}
\usepackage{ulem}
\usepackage{soul}

\usepackage{fancyhdr}
\usepackage{hyperref}

\begin{document}

\noindent
\begin{minipage}{\textwidth}
    \centering
    \footnotesize Published version available at Physical Review Letters: ~\href{https://doi.org/10.1103/621t-vx16}{https://doi.org/10.1103/621t-vx16}
\end{minipage}
\vspace{1em}
\preprint{APS/123-QED}

\author{Melissa Yactayo}
\email{melissa.yactayo@unmsm.edu.pe}
\affiliation{Université de Lorraine, CNRS UMR 7198, Institut Jean Lamour, Nancy, France}
\affiliation{Universidad Nacional Mayor de San Marcos, P.O.-Box 14–0149, Lima 14, Peru}
\author{A. Pezo}
\affiliation{Laboratoire Albert Fert, CNRS, Thales, Université Paris-Saclay, France}
\affiliation{Institute for Solid State Physics, University of Tokyo, Kashiwa, 277-8581, Japan}
\author{J. L. Ampuero}
\affiliation{Université de Lorraine, CNRS UMR 7198, Institut Jean Lamour, Nancy, France}
\author{M. Tian}
\affiliation{Department of Physics, Key Laboratory of Low Dimensional Quantum structures and Quantum Control of Ministry of Education, Hunan Normal University, Changsha, China}
\affiliation{Key Laboratory of Quantum Materials and Devices of Ministry of Education, School of Physics, Southeast University, Nanjing, China}
\author{L. Badie}
\affiliation{Université de Lorraine, CNRS UMR 7198, Institut Jean Lamour, Nancy, France}
\author{J. Quispe-Marcatoma}
\affiliation{Universidad Nacional Mayor de San Marcos, P.O.-Box 14–0149, Lima 14, Peru}
\affiliation{Centro de Investigaciones Tecnológicas, Biomédicas y Medioambientales, Bellavista 07006, Callao, Peru}
\author{C. V. Landauro}
\affiliation{Universidad Nacional Mayor de San Marcos, P.O.-Box 14–0149, Lima 14, Peru}
\affiliation{Centro de Investigaciones Tecnológicas, Biomédicas y Medioambientales, Bellavista 07006, Callao, Peru}

\author{Y. Xu}
\affiliation{MIIT Key Laboratory of Spintronics, School of Integrated Circuit Science and Engineering, Beihang University, China}
\author{Sébastien Petit-Watelot}
\affiliation{Université de Lorraine, CNRS UMR 7198, Institut Jean Lamour, Nancy, France}

\author{Michel Hehn}
\affiliation{Université de Lorraine, CNRS UMR 7198, Institut Jean Lamour, Nancy, France}

\author{A. Fert}
\email{albert.fert@cnrs-thales.fr}
\affiliation{Laboratoire Albert Fert, CNRS, Thales, Université Paris-Saclay, France}

\author{J.-C. Rojas-S\'anchez}
\email{juan-carlos.rojas-sanchez@univ-lorraine.fr}
\altaffiliation{Temporary address: Universidad Nacional de Ingenieria, Lima-Peru}
\affiliation{Université de Lorraine, CNRS UMR 7198, Institut Jean Lamour, Nancy, France}

\title{Orbitronics: orbit-spin interconversions for long range orbital propagation enabled by spin current}
\date{\today}

\begin{abstract}
Orbitronics assumes long-range orbital currents, like spin currents in spintronics. However, recent theories and experiments challenge this view, showing rather local characters for orbital polarization and importance of orbit-spin conversions. In our work, angular momentum is created by ferromagnetic resonance or thermal gradients in Ni/(Pt)Ti/Au heterostructures and detected by conversion to charge current. The observed charge currents are independent of Ti thickness up to 60 nm, which is incompatible with orbital transport and Orbital Hall Effect in Ti. Instead, they depend on both Ti interfaces, evidencing spin-mediated transport in between after and before local orbit–spin interconversions.

\end{abstract}


\maketitle

The generation and manipulation of orbital angular momentum (OAM) have emerged as central pillars of orbitronics \cite{Bernevig2005,Kontani2009,Go2018,PhysRevB.106.104414}, offering a potential path toward low-power devices. At the heart of this field lies a fundamental, unresolved debate: the characteristic length scale of orbital transport. Initial experimental and theoretical frameworks suggest that orbital currents can propagate over macroscopic distances, implying that orbital accumulation diffusion is a bulk-mediated phenomenon analogous to spin accumulation diffusion \cite{Choi2023,Hayashi2023,Hayashi2024}. Moreover, theoretical predictions indicate that, in some light metals such as V, Zr, and especially Ti, the orbital Hall conductivity ($\sigma_{OH}$) can significantly exceed its spin Hall conductivity ($\sigma_{SH}$) \cite{Go2024, Tanaka2008}, thus that the orbital current induced by Orbital Hall Effect (OHE) should be larger than the spin current induced by Spin Hall Effect (SHE). In Ti, several reports have estimated remarkably long orbital diffusion lengths ($l_{of}$) of orbital currents in bulk metals, ranging from 47 to 61 nm \cite{Hayashi2023, Choi2023}, or smaller, such as 5.3 nm \cite{Sun2025}, 3.2 nm - 3.5 nm \cite{Santos2025,santos2026probing} or 4.6 nm \cite{Hayashi2024}. In contrast, recent theoretical advances \cite{Kelly2024, Valet2025,ValetPRL2025,Ning2025} suggest that OHE is extremely small in inversion and time reversal symmetric metallic lattices, with also Orbital Accumulation relaxing within a few atomic layers. On the experimental side, some results support this short-range trend \cite{Kumar2025, Liu2025}. This perspective shifts the paradigm of orbital production and long-range transport in bulk metals to another one combining several types of interfacial productions and local conversions between orbital and spin. It is important to note that while the propagation in bulk metals remains a subject of intense debate, interfacial conversion via the Orbital Edelstein Effect (OEE) and its inverse (IOEE) is well-established \cite{Sergio2023,Go2018,Park2011}. This interfacial mechanism is fundamentally analogous to its spin counterpart, the spin-to-charge conversion predicted and observed in spin-texture systems, such as Rashba interfaces and topological insulators \cite{Rojas-sanchezPRL2016,Rojas-sanchez2013_Ag-Bi, MihaiMiron2010}. This has been observed mainly at metals interfaced with oxides such as Cu/MgO \cite{Xu2024}, Cu/CuO$_x$\cite{Santos2024,Kim2023,Ding2020HarnessingOC}, but also Co/Al \cite{Nikolaev2024} where orbital texture is presented. 

Clarifying the generation and relaxation length of OAM is crucial for guiding the design of future orbitronics architectures. If, as our experiments suggest, OAM generation and propagation are inherently short-distance, the orbitronics framework inspired by conventional bulk spin-current transport in spintronics must be reconsidered. Instead, a mechanism centered on local (mainly interfacial) orbital productions followed, for long distance effects, by conversions from orbital to spin, magnetization transport by spin current before spin to orbital reconversion, enabling the observation of long-distance effects of orbital origin.

\begin{figure}[h]
\centering
\includegraphics[width=0.5\textwidth]{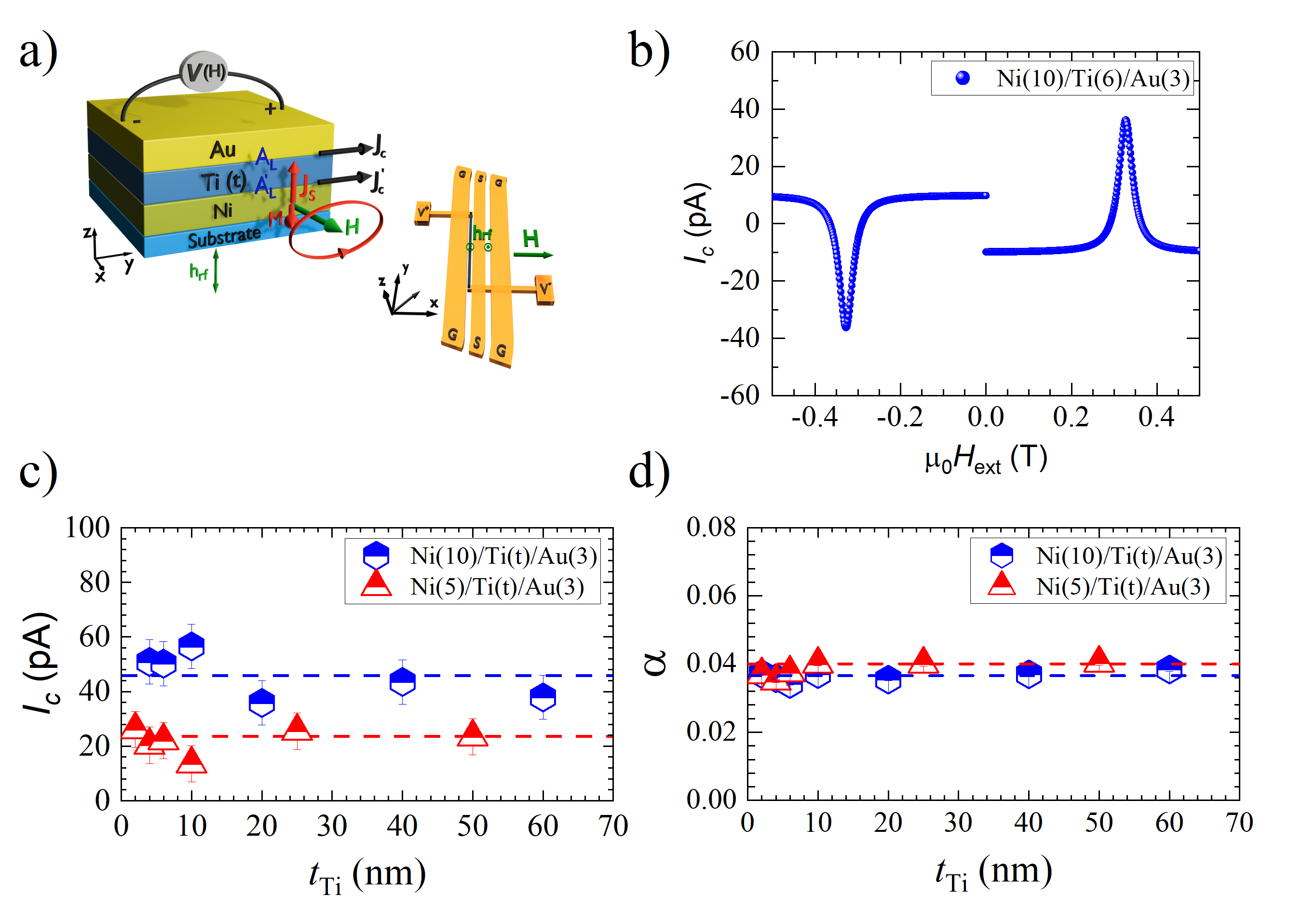}
\begin{quote}
\caption{SOP-FMR experiments. (a) Schematic of spin/orbital injection from Ni into Ti under S/OP-FMR and microwave antenna
geometry; as it will be shown spin and Orbital accumulations convert into a transverse charge current at the Ni/Ti and Ti/Au interfaces, $J_c^{'}$ and  $J_c$, respectively. 
(b) Symmetric odd component at 15 GHz and 15 dBm for Ni/Ti/Au 
(c) Charge current production $I_c$ for Ni(5 \& 10)/Ti(t)/Au(3): 
its independence from $t_{Ti}$ indicates an interface-dominated mechanism.
(d) Effective magnetic damping vs. $t_{Ti}$; whose flatness confirms the interfacial (not bulk) origin of angular momentum relaxation in Ti.} \label{Fig1}
\end{quote}
\end{figure}

In this Letter, we address this controversy by experiments with Spin/Orbital Polarization production from a Ni layer by Spin/Orbital pumping ferromagnetic resonance (SOP-FMR) \cite{rojas2014spin,Tserkovnyak2002, Rojas-Sanchez2013} or by thermal gradient (Spin/Orbital Seebeck, or SOSE, as in Spin Seebeck experiments \cite{Rezende2014,Anadon2022,Palin2023}). Ni is a well-known magnetic material to produce orbital accumulation \cite{GO-Ni-2025,Pezo2025}. In our samples, a Ni layer is below a Ti layer or several types of Ti-based heterostructures. We detect the voltage signal due to spin or orbit to charge conversions in the Ti-based heterostructures at the ferromagnetic resonance condition in SOP-FMR, Fig. \ref{Fig1}a. Spin pumping FMR is a well-established technique to study and quantify the efficiency of spin-to-charge interconversion \cite{rojas2014spin,Tserkovnyak2002, Rojas-Sanchez2013} and can be applied to quantify the orbital-to-charge interconversion. Our results demonstrate that the voltage signals obtained in both experiments are independent of the Ti-thickness ($t_{\text{Ti}}$), providing conclusive evidence of the interfacial nature of the conversion from spin and orbit to charge which is in agreement with our DFT theoretical calculations of interfacial conversions at different interfaces.

The samples were grown by DC magnetron sputtering and then patterned by standard UV lithography (See S5 supplemental material, SM). The width $\rm w$ of the slabs for both experiments is 10 $\mu$m. The series of samples to study Ti thickness dependence, are //Ni(5 \& 10)/Ti($t$)/Au(3) and //Ni(5)/Pt(2)/Ti($t$)/Au(3), where $t_{Ti}$ range from 2 to 60 nm. The number in brackets stand for the thickness in nm, and $''//''$ represents the position of the Si/SiO$_2$(500) substrate. 
The Au capping layer was deposited on all samples to prevent oxidation. This was distinguished by chemical analysis via Electron energy-loss spectroscopy and High-resolution Transmission electron microscopy (HRTEM). Structural characterization by X-ray Diffraction and HRTEM reveals that our samples are textured in the grown direction along (111) for FCC Ni, (002) for HCP Ti, and (111) for FCC Au (see details in S1 and S2 in SM). In SOP-FMR experiments, 10 nm Ni layer can still contribute to unwanted rectifying effects that mimic SOP-FMR voltage, when the radio frequency field, $h_{rf}$, is parallel to the sample interface. However these rectifications effects are strongly reduced when the FM thickness layer is reduced. Moreover, when $h_{rf}$ is perpendicular to the sample interface, along $z$ in Fig. \ref{Fig1}a, this spurious rectification effect is also suppressed \cite{Yactayo2026} (See S4 in SM). Consequently, in Fig. \ref{Fig1} we show results when $h_{rf}$ is along $z$. 
Furthermore, measurements of the total sheet conductance versus $t_{\text{Ti}}$ reveal a linear scaling across the entire thickness range, confirming the uniform physical properties of the Ti layers (See S6 in SM). \\

Figure \ref{Fig1}b shows the charge current production due to SOP-FMR effect, $I_c=V^{odd}_{sym}/R$ versus the applied DC magnetic field of a representative sample Ni(10)/Ti(6)/Au(3). Here, $V^{odd}_{sym}$ represents the symmetric Lorentzian component isolated through the antisymmetrization protocol, $V_{\text{S/O}} = \frac{V_{\text{sym}}(H>0) - V_{\text{sym}}(H<0)}{2}$, $R$ is the resistance of the full stack of the spin pumping slab. As discussed later in the manuscript, this conversion signal is significantly enhanced upon inserting a Pt layer between Ni and Ti to introduce conversions between orbit and spin. However, the initial discussion focuses on heterostructures without $\text{Pt}$ to clearly establish the purely interfacial nature of the primary mechanism.

Figure \ref{Fig1}c shows the results of Ti-thickness dependence on $I_c$ in Ni(5)/Ti(t)/Au(3) and Ni(10)/Ti(t)/Au(3). The results for the two series show that $I_c$ is independent of the Ti layer thickness, indicating that the underlying mechanisms of the conversions from spin and orbit to charge are not the bulk mechanisms of ISHE and IOHE. A negligible ISHE in Ti is not surprising for Ti due to its very small spin-orbit coupling and spin Hall conductivity \cite{Go2024}. The important result is the absence of significant contribution for IOHE in Ti, despite the expected orbital production by pumping from Ni. We ascribe this result to a very short orbital diffusion length in Ti preventing the penetration of orbital polarization beyond an interfacial zone of a few atomic distances and any conversion to charge inside Ti, as confirmed later by the analysis of interfacial conversions from orbital to charge current.

To also confirm the absence of orbit and spin absorption of the Ti layer, the magnetic damping constant ($\alpha$) in Ni was extracted for each heterostructure from broadband ferromagnetic resonance measurements. As shown in Fig. \ref{Fig1}d, $\alpha$ remains invariant across all sample series. This absence of thickness dependence indicates the absence of any damping enhancement due to spin or orbital conversion to charge (i.e absence of spin and orbital absorption) in Ti, confirming the absence of orbital penetration and conversion to charge that we deduced from the absence of variation of the charge current with the Ti thickness. This absence of thickness dependence also indicates that the intrinsic magnetic properties of the Ni layer, characterized by a saturation magnetization $M_s = 428$ kA/m and $M_s = 395$ kA/m, for Ni(10) and Ni(5), respectively, are preserved in samples of the same series (See S3 in SM). 

To further corroborate these findings, we performed the magnetothermal measurements described in End Matter. They confirm the SOP-FMR results: no orbital propagation beyond a few atomic distances from the Ni/Ti interface and only interfacial production of thermo-current.

If an IOHE process hold, the conversion signal would scale with $t_{\text{Ti}}$ through the characteristic diffusive model result, $\tanh(t/2l)$, which was originally established for spin transport \cite{Fert_Valet1993, rojas2014spin} and now extended to orbitronics \cite{Hayashi2024,Sun2025,Santos2025}. Following this model for orbital-to-charge conversion in the diffusive limit, $I_c$ and $i_{thermo}$ signal follow a $\tanh(t_{\text{Ti}}/2l_{\text{of}})$ increasing until reaching saturation at $t_{\text{Ti}} \approx 4l_{\text{of}}$, where $l_{\text{of}}$ would be the orbital diffusion length of Ti. In our experiments, $I_c$ (Fig. \ref{Fig1}c) shows a plateau starting at $t_{\text{Ti}} \approx 4$ nm (or below) lead us to estimate $l_{\text{of}} \approx 1$ nm (or smaller), indicating that orbital accumulation is confined inside the interfacial region and that orbital propagation does not go beyond this zone. Consequently, our data point to an interfacial conversion mechanism. These results stand in sharp contrast to previous reports on Ti \cite{Hayashi2024, Choi2023}. In our work, after having validated the structural, magnetic and electronic  quality of our heterostructures (See S2, S3 and S6 in SM), we will be able to provide an alternative interpretation of the orbital-to-charge conversion in high-quality pure metals.

As, in both spin-orbital pumping and magnetothermal experiments, the signal cannot be ascribed to IOHE and ISHE in Ti, a fundamental question arises regarding the origin of the measured charge current: where does the conversion of spin and orbital accumulation into $I_c$ occur, and what are the governing mechanisms? Our experimental evidence points to a purely interfacial conversion and the second step of our work is to determine whether this process is localized at the primary Ni/Ti interface, the secondary Ti/X interface, or arises from contribution of both.

To investigate these interfacial contributions,
we studied now a series of heterostructures consisting of //Ni(5)/Ti(6 or 10)/X(4) multilayers, and the X capping layer being Al, W, Pt, and Au. 

\begin{figure}[h!]
\centering
\includegraphics[width=0.4\textwidth]{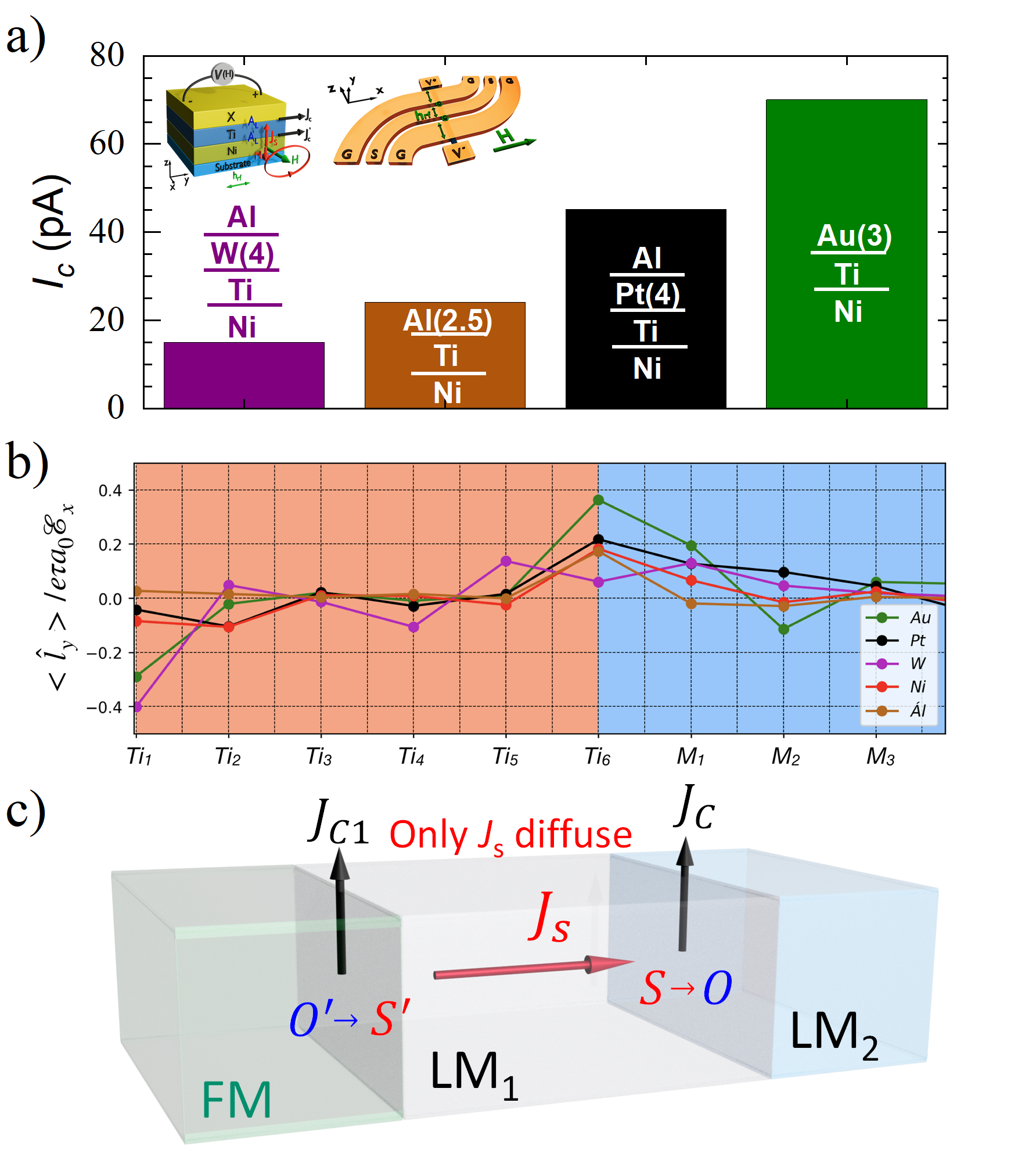}
\begin{quote}
\caption{Orbital to charge conversions at the Ti top interface. a)
Total charge current ($I_c$) in Ni/Ti/X heterostructures (X = Al, W, Pt, Au), showing how the capping layer modulates overall efficiency (10 GHz, 15 dBm).
Inset: experimental S/O-FMR configuration with microwave antenna. b) DFT calculations of field-induced orbital polarization ($l_y$) at Ti/(FM \& NM) interfaces, giving the Edelstein orbital-to-charge conversion efficiency. Predicted trend matches experiment: Ti/Au $>$ Ti/Pt $>$ Ti/Al $>$ Ti/W. c) Schematic of the double-interconversion mechanism in FM/LM$_1$/LM$_2$ heterostructures. Part of the spin and orbital accumulations generated at the FM/LM$_1$ interface convertes into a charge current $J_{c1}$. The remaining orbital accumulation, unable to propagated in LM$_1$, converts to spin polarization (O to S) and diffuses as spin current $J_s$ to LM$_2$. There, $J_s$ induces a spin accumulation and local S-to-O conversion, and the resulting orbital accumulation is converted into charge current, $J_c$, via the inverse orbital Rashba–Edelstein effect (IOREE). This O loading-unloading process requires  LM$_1$$'\text{s}$ spin diffusion lenght ($l_{sf}$) to be at least as large as its thickness- for Ti, our experiments (SM,S4) indicate $l_{sf} > 80$ nm.} \label{Fig3}
\end{quote}
\end{figure}

As evidenced by the results in Fig. \ref{Fig3}a, $I_c$ signal depends significantly on the capping layer. This clear dependence demonstrates that the overall signal does not originate only from the Ni/Ti interface but depends significantly on also the top Ti/X interface. $I_c$ is enhanced in the presence of Au and Pt capping layer but reduced by W, and Al. Notably, the largest signal is achieved in the Ni/Ti/Au configuration.

To elucidate if the modulation at the top interface comes from spin or orbital texture, i.e Spin Rashba-Edelstein effect (SREE) or  Orbital Rashba-Edelstein effect (OREE), we performed first-principles calculations. Spin ($\hat{s}_y$) and  orbital ($\hat{l}_y$)  polarization along $y$  induced by an in-plane field ($E_x$) were calculated for Ti interfaced with Al, Ni, W, Pt, and Au. Calculations reveal that interfacial charge-to-spin conversion (SREE) is an order of magnitude weaker than charge-to-orbital conversion (OREE) (see S8 in SM \cite{PhysRevB.111.L140405,apezol_CoAl,zerothi_sisl,Liu2024,ovalle_2023,Krishnia:2024,Blochl:1994,Kresse:1999,Kresse:1996a,Kresse:1996b,siesta_method,Perdew:2008,Perdew:2009,siesta_on-site_soc,tm_pseudopotentials,zerothi_sisl}). This confirms that the experimental signal at the top Ti interface arises mainly from orbital-to-charge conversion (IOREE), most efficient at Ti/Au (Fig. \ref{Fig3}b). 
The signal ordering Ti/Au $>$ Ti/Pt $>$ Ti/Al $>$ Ti/W further supports dominant orbital effects, matching the theoretical Edelstein-effects ratios $l_y/E_x$, from the variation of $\frac{\hbar\braket{\hat{l}_y}}{eE_xa_0\tau}$, shown in Fig. \ref{Fig3}b.

Furthermore, our experimental and theoretical results clarify whether the signal is primarily driven by conversion at the bottom or top interfaces of the Ti layer. According to the results ($l_y/E_x$) in Fig.~\ref{Fig3}b, the conversion exhibits the same positive sign for all interfaces Ti/X and Ti/Ni. In the specific case of Ti/Ni and Ti/Au interface, this symmetry leads to opposite signal signs upon reversing the stacking order (i.e., Ni/Ti and Ti/Au yields opposites signs). Remarkably, the positive polarity of our measured current signals matches the theoretical prediction for orbit-to-charge conversion at the Ti/Au and others top interfaces. Consequently, we conclude that the signals observed in our spin-orbital pumping experiments originate from the predominance of inverse orbital Rashba-Edelstein effects (IOREE) at the top Ti interfaces (with largest effects for Ti/Au) on opposite effects at bottom of Ti.

If we have IOREE conversion at the top interface, how the orbital polarization traverses Ti? We begin by highlighting the results of our additional measurements on Fe/Ti/Pt structures showing that the spin diffusion length in Ti is longer than, at least, 80 nm, a value significantly larger than in previous reported values (e.g., $13\text{ nm}$ in Ref.\cite{du2014PRB}). In the framework of Valet-Fert theory, the spin diffusion length is related by spin-flip scattering at defects and impurities. Therefore, our long spin diffusion length is consistent with the high structural quality and uniformity of our Ti layers (see SM, S1 and S2), as further evidenced above by the invariance of the Ni damping with Ti thickness, thereby allowing a spin current to traverse $60\text{ nm}$ of Ti without significant attenuation (see SM, Sec.~S4.A). It justifies our interpretation of orbital effects at top interface of Ti by the following multistep transport process. After injection of orbital and spin angular momentum from the Ni layer, the orbital and spin accumulation at the Ni/Ti interface are 1) partially converted to charge current ($J_{c1}$)
along the interface by IOREE and ISREE, and 2) there is local conversion of some part of the orbital accumulation into a spin accumulation (O $\rightarrow$ S) and the resulting generation of a part of spin current ($J_s$) traversing Ti and reaching the top Ti/X interfaces. There, after some reconversion from spin into orbital accumulation (S $\rightarrow$ O), the orbital and spin accumulations are (partly) converted into a charge current by IOREE or ISREE at the top interface, especially at Ti/Au interface (the inverse spin Hall effect (ISHE) of Ti is too weak to generate a measurable additional signal). Our proposed multi-step mechanism is schematized in Fig. \ref{Fig3}c, a mechanism for non-local orbital effects despite ultra-short orbital diffusion.

\begin{figure}[h!]
\centering
\includegraphics[width=0.5\textwidth]{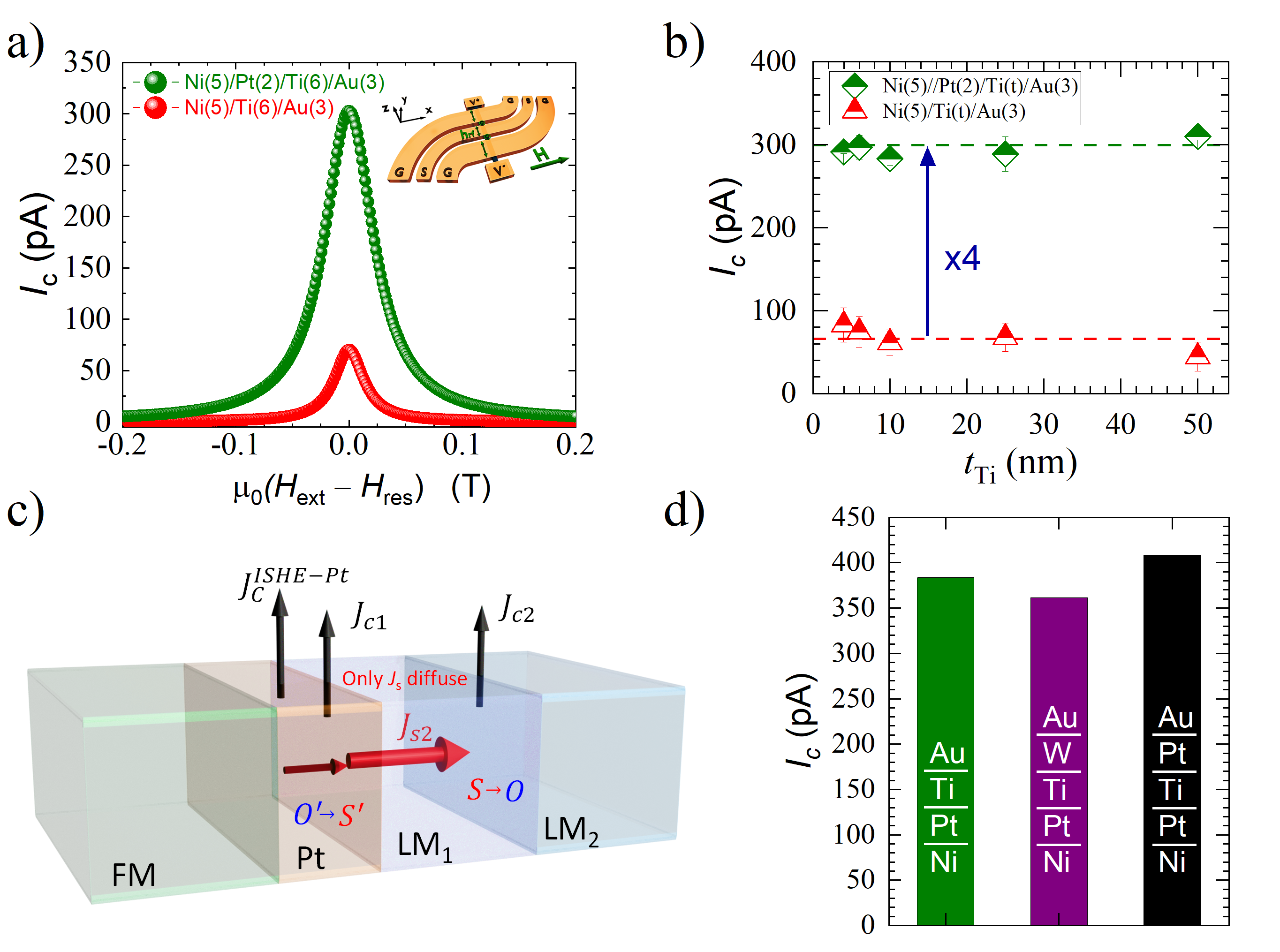}
\begin{quote}
\caption{Signal enhancement after addition a 2 nm Pt layer. a) Symmetrical odd component of $I_c$ vs field for Ni/Ti/Au and Ni/Pt/Ti/Au (10 GHz, 15 dBm). The insertion of Pt layer strongly enhances the signal. Inset: Microwave antenna geometry. b) $I_c$ is independent of Ti thickness for both stacks, with (green) and without (red) Pt.
c) Multiple-interconversion mechanism in FM/HM/LM$_1$/LM$_2$ heterostructures: orbital accumulation from Ni is efficiently converted by Pt into spin accumulation, generating a large spin current $J_s$ that carries spin magnetization to LM$_2$, where it reconverts into orbital magnetization at LM$_1$/LM$_2$ interface (e.g. Ti/Au, Ti/Pt). d) $I_c$ for //Ni(5)/Pt(2)/Ti(10)/X(4) separate deposition batch, showing dependence on capping layer X.} 
\label{Fig4}

\end{quote}
\end{figure}

To further validate the proposed mechanism, and confirm the central role of the conversion from orbital to spin accumulation, we inserted 2 nm Pt layer between the Ni and Ti, //Ni(5)/Pt(2)/Ti(t)/Au(3), where $t_{Ti}$ range from 2 to 50 nm. Strong spin-orbit coupling (SOC) materials such as Pt or rare earths are known to efficiently convert orbital accumulation into spin accumulations($O \rightarrow S$) \cite{Santos2023-Pt, Costa2025,Gambardella_rare_earths_2022,Ding_rare_2024}. Theoretically, these conversions are now described as local conversions between orbital and spin accumulations and not between orbital and spin currents \cite{Ning2025}, a local character consistent with our interpretations. The conversion enhancement by Pt is clearly reflected in Fig.\ref{Fig4}a, where $I_c$ ($I_c \approx 70$ pA) in the Ni(5)/Ti($t$)/Au(3) heterostructures increase significantly to $I_c \approx 300$ pA upon Pt insertion. This nearly fourfold increase is consistently observed across the whole Ti thickness range (Fig.\ref{Fig4}b) in the Ni/Ti/Au and Ni/Pt/Ti/Au series. Remarkably, even with the insertion of 2 nm of Pt layer, the enhanced charge current remains independent of the Ti thickness, further corroborating the purely interfacial nature of the conversion mechanism.\\

For a more quantitative interpretation we have also estimated other contributions to the enhancement of the signal to 300 pA after insertion of Pt. The bulk inverse Spin Hall Effect (ISHE) in Pt accounts for only 170 pA, calculated from total $I_c$ in //Ni(5)/Pt(10) bilayer measured under the same conditions (See S4.E in SM \cite{JC-rojasSanchez-PRL2014}). In addition, the insertion of Pt introduces additional Ni/Pt and Pt/Ti interfaces interfaces that could affect spin transparency or spin memory loss, and possibly could contribute to the total signal, but these local interface effects cannot reasonably account for the fourfold signal enhancement. Finally, the enhancement of the current by Pt can be essentially ascribed to the enhancement of the conversion between O and S by Pt, as already reported with other heavy elements in other experiments \cite{Santos2023-Pt,Costa2025} and theoretical works \cite{Ning2025}. When the enhancement of the signal by the insertion of Pt is taken into account, the picture of Fig. \ref{Fig3}c becomes the picture of Fig. \ref{Fig4}c in which the O to S conversion shown in the Pt gives rise to a larger spin current in LM$_1$ (stronger red arrow than in Fig. \ref{Fig3}c.)

To see again the dependence on top interfaces with Pt insertion at bottom, we show the results on Ni(5)/Pt(2)/Ti(10)/X(4)/Au(3), ($X = \text{Pt, W}$), Fig. \ref{Fig4}d.
At this stage, and without additional measurement on many samples, it is impossible to go more deeply into the separation of the different conversions to charge at different interfaces. The only clear information that we can derive is that, in contrast with the large enhancement (x4) brought by the O to S conversion when Pt is at the bottom, placing Pt and W heavy metals on top do not introduce large changes. Probably, it confirms the general theoretical prediction of weaker S to O conversion compared to O to S in relation to the much smaller orbital diffusivity for orbit compared to spin, Fig.2 in \cite{Ning2025}. This can be described as a non-reciprocity of the conversions between spin and orbit.

One of the results of our work is the very short orbital diffusion length in Ti, $l_{of} \approx 1$ nm (or smalller), not very far from some results in recent works such as 2 nm in V \cite{aguilar2025orbital}, 1.7 nm in Zr \cite{Kumar_Zr_2025}, 5.3 nm in Ti \cite{Sun2025} but much shorter than many results as, for example 61 nm in Ti \cite{Choi2023}. We discuss the possible origin of this dispersion in S4.C, SM, including the possible enhancement of the orbital diffusion length by experimental factors, as, for example, underestimation of rectification effects \cite{Stiles2025,Liu2025,Ampuero_FePt,Yactayo2026} or by disorder proposed by \cite{Belashchenko2023,manchon2025multipolarorbitalrelaxationt2g,Ning2025,Liu2025-disorder}.
Finally, we can also mention our quantitative results on the global spin and orbit  conversion efficiency length \cite{Rojas-Sanchez2019} in our SOP-FMR experiments, obtaining $\lambda = 19\text{ pm}$ for Ni/Ti/Au, as detailed in S4.B in SM, a relatively small value showing the interfacial effects can be significant even with modest interfacial Edelstein effects.

In conclusion, we have first demonstrated that the orbital accumulation generated by SOP-FMR or thermal gradient at the bottom interface of Ti layers does not penetrate beyond 1 nm in quasi-epitaxial Ti layer, the corresponding very short orbital relaxation length obtained in this work provides a direct experimental confirmation of recent theoretical predictions for inversion and time symmetric metallic lattices.
The conversion of orbit and spin into charge current depends on the choice of X at the Ti/X top interface of Ti and can be ascribed to predominant interfacial effects (IOREE and ISREE) at this interface. A signature of predominant orbital contributions (IOREE) comes from the agreement with the respective ab-initio calculations for different Ti/X interfaces. The importance of orbital effects at long distance (up to 60 nm of Ti in our samples) can be explained by the multistep mechanism scenario that we have proposed and tested: in the notation in Fig. \ref{Fig3} and Fig. \ref{Fig4}, Orbital (O) and spin (S) angular momentum are pumped from the FM layer, generating O and S accumulations at the FM/LM$_1$ interface, the O accumulation being partly converted into current along the FM/LM$_1$ interface and partly converted from O to S to contribute to the spin current diffusing through the LM$_1$ layer (loading step). At the top interface (LM$_1$/LM$_2$), S is converted back into O (unloading), and O can be subsequently converted into a charge current. Our results clarify the effects of the spin–orbit–charge interconversions in long-distance effects in orbitronics and pave the way for the design of future orbitronics applications.

\section*{Data Availability}
The data that support the findings of this article are publicly available \cite{Orbital_Ni_Ti_dataset_2026_zenodo}

\medskip
\textbf{Acknowledgements}\\ 
This work was funded by the ERC CoG project MAGNETALLIEN grant ID  101086807, the EU-H2020-RISE project Ultra Thin Magneto Thermal Sensoring ULTIMATE-I (Grant ID. 101007825), and the French National Research Agency (ANR) through the project “Lorraine Université d’Excellence” reference ANR-15-IDEX-0004-LUE. It was also partially supported by the ANR through the France 2030 government grants EMCOM (ANR-22-PEEL-0009), PEPR SPIN ANR-22-EXSP-0007 and ANR-22-EXSP-0009. Devices in the present study were patterned at Institut Jean Lamour's clean room facilities (MiNaLor). M. Yactayo is grateful to MSCA RISE ULTIMATE-I (Grant ID. 101007825) and CONCYTEC for the funding provided through the PROCIENCIA under Undergraduate and Postgraduate Theses in Science, Technology and Technological Innovation 2025-01 program (PE501098380-2025). A.P. was supported by the Japan Science and Technology Agency (JST) ASPIRE Program No. JPMJAP2410, the ANR 'ORION' project, grant ANR-20-CE30-0022-01 of the French Agence Nationale de la Recherche, ANR-22-CE30-0026 'DYNTOP' project. High Performance Computing resources were provided by the EXPLOR centre hosted by the University of Lorraine ( Project: 2026M4XXX4158).
\clearpage

\bibliography{biblio-Ni-Ti}


\clearpage 

\section*{End Matter}
\begin{figure}[h!]
\centering
\includegraphics[width=\columnwidth]{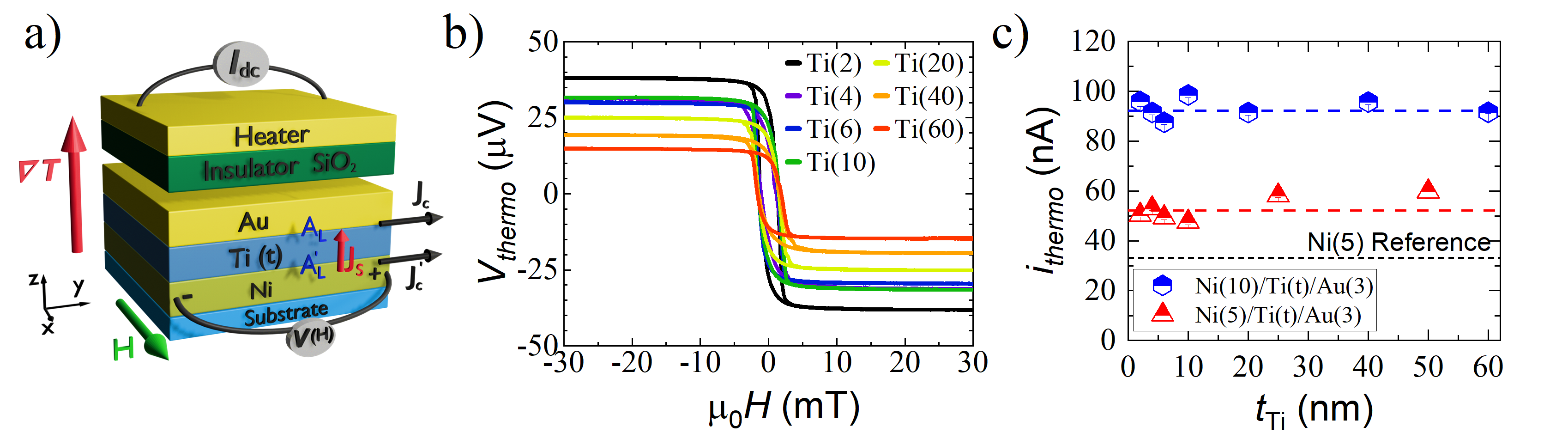}
\caption{Thermal generation and conversion of spin and orbital accumulation. (a) Schematic of the spin and orbital Seebeck effects. A vertical thermal gradient ($\nabla T$) generates a spin or orbital accumulation, which is injected from the Ni layer into the detection layers 
(b) Raw data of the voltage detected as a function of external magnetic field in Ni(10)/Ti(t)/Au(3) for a fixed heater power (100 mA). 
(c) Total thermocurrent (including the anomalous Nernst effect contribution) for Ni(10)/Ti($t$)/Au(3) and Ni(5)/Ti($t$)/Au(3) series. The horizontal dotted line represents the ANE of the $5\text{nm}$ Ni layer ($\text{Al}(2.5)/\text{Ni}(5)/\text{Al}(2.5)$ reference).
The no variation of $i_{thermo}$ with respect to $t_{\text{Ti}}$ suggests that orbital-to-charge conversion does not occur within the Ti bulk but is instead dominated by interfacial contributions.
} \label{Fig2}
\end{figure}

\textit{Appendix: Magnetothermal measurements} 
To further corroborate the findings presented in the main text by SOP-FMR (Fig.~\ref{Fig1}a--c), we performed magnetothermal measurements as illustrated in the experimental scheme in Fig.~\ref{Fig2}(a).
Unlike the microwave-driven resonance in SOP-FMR experiments (Fig. \ref{Fig1}a) the SOSE is governed by a thermal gradient, $\nabla T$ \cite{Rezende2014,Anadon2022,Palin2023}. Fig.~\ref{Fig2}b shows the raw data voltage as a function of the external magnetic field for the Ni(10)/Ti(t$_{Ti}$)/Au(3) series under a fixed heater current of 100 mA. From these curves, we extract the thermovoltage contributions at positive and negative saturation fields, $V_{\text{thermo}}^{H>0}$ and $V_{\text{thermo}}^{H<0}$, respectively. These values allow us to determine the thermo-current ($i_{thermo}$), defined as $i_{\text{thermo}} =| (V_{\text{thermo}}^{H>0} - V_{\text{thermo}}^{H<0})|/R$, where $R$ is the four-probe resistance of the full stack of the strip measure in a length of $L= 200$ $\mu$m (See S5 in SM). Following this equation, the resulting $i_{\text{thermo}}$ for both the $\text{Ni}(5)$ and $\text{Ni}(10)$ series are presented in Fig.~\ref{Fig2}c. The vertical offset between the two series is attributed to the anomalous Nernst effect (ANE), which scales with the ferromagnetic layer thickness. 
To benchmark this contribution, we measured an $\text{Al}(2.5)/\text{Ni}(5)/\text{Al}(2.5)$ control sample, which yields an intrinsic ANE of $\approx 33\text{ nA}$ generated only by the $5\text{-nm}$ Ni layer, lower than the average value of $\sim 51\text{ nA}$ obtained across the $\text{Ni}(5)/\text{Ti}(t)/\text{Au}(3)$ series. Because the Ni layer thickness is identical across this entire series, the ANE acts purely as a constant baseline offset of $\sim 33\text{ nA}$, thereby confirming that it does not obscure the interfacial nature of the inverse orbital Edelstein effect (IOREE) at the $\text{Ti}/\text{Au}$ interface.

Crucially, however, $i_{\text{thermo}}$ remains invariant with respect to $t_{\text{Ti}}$ across all series. This independence rules out bulk orbital-to-charge conversion within the Ti layer, confirming again a predominantly interfacial origin for the conversion mechanisms of orbital polarization.
\end{document}